\documentclass[letterpaper,english,reprint, aps]{revtex4-1}
\usepackage[T1]{fontenc}
\usepackage[latin9]{inputenc}
\usepackage{mathtools}
\usepackage{amsmath}
\usepackage{amssymb}
\usepackage{graphicx}
\usepackage{esint}

\makeatletter

\pdfpageheight\paperheight
\pdfpagewidth\paperwidth

\usepackage{xcolor}
\usepackage[unicode=true,
 bookmarks=false,
 breaklinks=false,pdfborder={0 0 1},backref=section,colorlinks=true]
 {hyperref}
\hypersetup{
 linkcolor=purple,urlcolor=blue,citecolor=blue}

\makeatother

\usepackage{babel}
\begin{document}
\title{Thin-Shell Wormhole Satisfying Energy Conditions}
\author{S. Danial Forghani}
\email{danial.forghani@final.edu.tr}

\affiliation{Faculty of Engineering, Final International University, Kyrenia, North
Cyprus via Mersin 10, Turkey}
\author{S. Habib Mazharimousavi}
\email{habib.mazhari@emu.edu.tr}

\affiliation{Department of Physics, Faculty of Arts and Sciences, Eastern Mediterranean
University, Famagusta, North Cyprus via Mersin 10, Turkey}
\author{Mustafa Halilsoy}
\email{mustafa.halilsoy@emu.edu.tr}

\affiliation{Department of Physics, Faculty of Arts and Sciences, Eastern Mediterranean
University, Famagusta, North Cyprus via Mersin 10, Turkey~\\
}
\date{\today}
\begin{abstract}
The quartic self-interacting conformal scalar field is used to construct
a thin-shell wormhole satisfying all energy conditions. Accompanying
the scalar field is the extremal Reissner-Nordström black hole with
a positive cosmological constant. New junction conditions apt for
the higher-order terms are introduced in the Gaussian normal coordinates.
Our approach may provide a guideline towards getting rid of exotic
matter in TSWs.
\end{abstract}
\maketitle

\section{\label{section1}Introduction}

Black holes with the hair of conformal scalar field entered into physics
literature with the Einstein-Maxwell-Scalar (EMS) solutions, given
by Bochorova, Bronikov and Melnikov \citep{Bocharova1}, and later
independently by Bekenstein \citep{Bekenstein1}. These works will
be referred together as BBMB. In obtaining this solution, the extremal
Reissner-Nordström (ERN) solution was used as base on which the scalar
field was added. The scalar field itself turned out to be singular
at the horizon whereas the spacetime was regular, as it should be.
The interest in such a solution was due to the fact that a test scalar
field, in contrast to the coupled scalar field, did not feel the singularity
so that the physics went normal at the horizon. Conformal scalar field
is important because it is related to the scaling of spacetime coordinates.
The history of such scaling, in connection with scalar fields and
their use in field theory, goes back to the idea of Lifshitz . The
spinless scalar interaction between objects naturally provides the
simplest type of interaction in both classical and quantum field theory.
Global scaling of coordinates may be valid in a spacetime leading
to certain conserved quantities. More importantly, the local scaling
which amounts to $g_{\mu\nu}\rightarrow\varOmega^{2}g_{\mu\nu}$,
and scalar field $\psi\rightarrow\varOmega^{-1}\psi$, change the
equation satisfied by the scalar field $\psi$, so that the revised
version becomes of the form $\nabla^{2}\psi-\frac{1}{6}R\psi=0$,
involving the scalar curvature of the spacetime. Further extensions
with conformal scalar of black hole solutions follow by taking into
account the higher order coupling of scalar fields higher than the
quadratic ones. One such solution with a positive cosmological constant
and conformal scalar fields was given by Mart\'{\i}nez, Troncoso and
Zanelli (MTZ solution) \citep{Martinez1}. Their solution was also
built on the ERN solution with positive cosmological constant, and
a static electric charge. However, as the BBMB solution introduced
a new scalar hair, the MTZ solution did not do so. It happens that
the scalar hair is expressed in terms of mass and charge. Furthermore,
the scalar field diverges on the event horizon as in the case of BBMB.
Yet, the latter is an interesting black hole solution covering a positive
cosmological constant, electric charge and a conformal scalar field
of fourth order. An interesting property of MTZ is that it satisfies
the energy conditions such as the strong and the dominant energy conditions
(SEC and DEC). This latter property attracted our interest and motivated
us to construct a thin-shell wormhole (TSW) that may satisfy physical
energy conditions without need of exotic matter. From this token,
we adopt the model of MTZ with minor rearrangements and apply the
method of TSW developed by Visser \citep{Visser1}. Our primary task
toward this goal is to derive new junction conditions in this new
theory of gravity. It is well-known that in Einstein's general relativity
such junction conditions were developed first by Israel \citep{Israel1}.
For new coupling terms in the Lagrangian, it is crucial that new junction
conditions are used. Previously, new junction conditions have been
developed for different theories \citep{Davis1,Reina1,Senovilla}
and in the same line of thought we do the same in the present theory.
We employ the Gaussian normal coordinate system and find the extrinsic
curvature within the set of Gauss-Codazzi equations. Although the
equations are tedious, we impose the conditions that upon crossing
the thin-shell metric functions are continuous while their normal
derivatives may admit discontinuity. We recall that these adopted
conditions are the standard ones used in physics in general. Upon
this approach, all equations on the surface reduce to simple conditions
that the problem becomes tractable enough to construct TSWs. The physical
finding is that above certain range of parameters such TSW admits
physically satisfactory energy conditions. As a result of considering
a quartic coupling term for a conformal scalar field with ERN and
positive cosmological constant, we construct a physical TSW, free
of exotic matter. Further models can be employed with new and higher
order couplings, provided the junction conditions are developed separately
in each case.

The order of the letter is as the following. In section \ref{section2}
we have developed the junction conditions apt for the gravity theory
considered here. In section \ref{section3} we construct a symmetric
TSW and study the situations under which the TSW satisfies the known
energy conditions. Finally, we conclude in section \ref{section4}. 

\section{\label{section2}Junction Conditions}

The action for a conformal massless scalar field $\psi$ coupled to
the cosmological Einstein-Hilbert density and other fields $\mathcal{L}_{M}$
is given by \citep{Martinez1}
\begin{multline}
I=\frac{1}{2\kappa}\int\sqrt{-g}d^{4}x\left[R-2\Lambda\left(1-\alpha^{2}\psi^{4}\right)\right.\\
\left.-\alpha\left(\psi^{2}R+6\partial_{\mu}\psi\partial^{\mu}\psi\right)+2\kappa\mathcal{L}_{M}\right],\label{action}
\end{multline}
in which $\kappa=8\pi G/c^{4}$ is the Einstein constant, $\Lambda$
is the cosmological constant, and $\alpha$ is a coupling constant.
Here we choose $c=1$ by convention. To obtain MTZ solution \citep{Martinez1},
the existence of the quartic interaction term to the cosmological
expression is inevitable. It is also noteworthy to point out that
the above Lagrangian density (without the matter field) is in fact
a special case of Horndeski theory \citep{Horndeski1} with $K\left(\psi,X\right)=\left(1/\kappa\right)\left[6\alpha X-\Lambda\left(1-\alpha^{2}\psi^{4}\right)\right]$,
$G_{4}\left(\psi,X\right)=\left(1/2\kappa\right)\left(1-\alpha\psi^{2}\right)$,
and $G_{3}\left(\psi,X\right)=G_{5}\left(\psi,X\right)=0$ \citep{Ganguly1}.
Varying this action with respect to the scalar field $\psi$ leads
to the conformal equation
\begin{equation}
\nabla^{2}\psi-\frac{1}{6}\psi R+\frac{2}{3}\alpha\Lambda\psi^{3}=0,\label{conformal equation}
\end{equation}
whereas variation with respect to the metric tensor $g_{\mu\nu}$
yields the field equation
\begin{multline}
G_{\mu\nu}\left(1-\alpha\psi^{2}\right)+\Lambda g_{\mu\nu}\left(1-\alpha^{2}\psi^{4}\right)=\\
\left[6\alpha\partial_{\mu}\psi\partial_{\nu}\psi-3\alpha\left(\partial\psi\right)^{2}g_{\mu\nu}\right.\\
\left.-\alpha\nabla_{\mu}\nabla_{\nu}\left(\psi^{2}\right)+\alpha\nabla^{2}\left(\psi^{2}\right)g_{\mu\nu}+\kappa T_{\mu\nu}\right],\label{field equation}
\end{multline}
in which $G_{\mu\nu}=R_{\mu\nu}-\frac{1}{2}Rg_{\mu\nu}$ is the Einstein
tensor and $T_{\mu\nu}$ is the energy-momentum tensor due to the
existing sources. Actually, the name conformal scalar field is due
to the fact that Eq. \eqref{conformal equation} is invariant under
the conformal transformation $g_{\mu\nu}\rightarrow\varOmega^{2}g_{\mu\nu}$
and $\psi\rightarrow\varOmega^{-1}\psi$, where $\varOmega$ is an
arbitrary differentiable function. However, since the Einstein-Hilbert
term is not invariant under such a transformation, the action in whole
is not conformal. By contracting Eq. \eqref{field equation} and utilizing
Eq. \eqref{conformal equation}, one can easily show that
\begin{equation}
R=4\Lambda-\kappa T,\label{Ricci scalar}
\end{equation}
which implies $R=4\Lambda=const.$ for a cosmological action without
a matter field, or a trace-free energy-momentum tensor (such as the
one of the electromagnetic field). Simultaneous consideration of Eqs.
\eqref{conformal equation}, \eqref{field equation}, and \eqref{Ricci scalar}
leads to an alternative form for Eq. \eqref{field equation}, i.e.
\begin{multline}
\left[R_{\mu\nu}-\Lambda g_{\mu\nu}\left(1+\frac{1}{3}\alpha^{2}\psi^{4}\right)\right]\left(1-\alpha\psi^{2}\right)+\alpha\Upsilon_{\mu\nu}\\
=\kappa\left[T_{\mu\nu}-\frac{1}{2}T\left(1-\frac{1}{3}\alpha\psi^{2}\right)g_{\mu\nu}\right],\label{field equoation 2}
\end{multline}
where we have defined
\begin{equation}
\Upsilon_{\mu\nu}=\nabla_{\mu}\nabla_{\nu}\left(\psi^{2}\right)+\left(\partial\psi\right)^{2}g_{\mu\nu}-6\partial_{\mu}\psi\partial_{\nu}\psi.\label{upsilon tensor}
\end{equation}
The Gauss-Codazzi equations for Ricci tensor elements in Gaussian
normal coordinates are given by
\begin{equation}
R_{wa}=-\nabla_{a}K+\nabla_{b}K_{a}^{b},\label{GC1}
\end{equation}
\begin{equation}
R_{ab}=\,^{\mathcal{H}}R_{ab}-\partial_{w}K_{ab}+2K_{a}^{c}K_{cb}-KK_{ab},\label{GC2}
\end{equation}
\begin{equation}
R_{ww}=-g^{ab}\partial_{w}K_{ab}+Tr\left(K^{2}\right),\label{GC3}
\end{equation}
for an $n$-dimensional hypersurface embedded in an $n+1$-dimensional
bulk spacetime. In this framework, $w$ is the spatial Gaussian normal
coordinate and the timelike hypersurface exists at $w=0$, so the
metric of the hypersurface could be depicted as
\begin{equation}
ds_{\mathcal{H}}^{2}=dw^{2}+g_{ab}dx^{a}dx^{b}.\label{induced metric}
\end{equation}
Furthermore, $\,^{\mathcal{H}}R_{ab}$ is the induced Ricci tensor
on the hypersurface, $K_{ab}\equiv\frac{1}{2}\partial_{w}g_{ab}$
and $K\equiv g^{ab}K_{ab}$ are the extrinsic curvature tensor and
the total curvature, respectively, and $Tr\left(K^{2}\right)\equiv K^{ab}K_{ab}$.
Upon direct substitution of the three Gauss-Codazzi equations into
Eq. \eqref{field equoation 2} with appropriate indices, after some
mathematical manipulation one could recover
\begin{multline}
\left(\nabla_{a}K-\nabla_{b}K_{a}^{b}\right)\left(1-\alpha\psi^{2}\right)\\
+\alpha\left[\partial_{w}\partial_{a}\left(\psi^{2}\right)-\varGamma_{wa}^{\mu}\partial_{\mu}\left(\psi^{2}\right)-6\partial_{w}\left(\psi\right)\partial_{a}\left(\psi\right)\right]=\\
\kappa\left\{ T_{iw}^{+}\varTheta\left(w\right)+T_{iw}^{-}\varTheta\left(-w\right)\right\} ,\label{eq:11}
\end{multline}
\begin{multline}
\left[\,^{\mathcal{H}}R_{ab}-\partial_{w}K_{ab}+2K_{a}^{c}K_{cb}-KK_{ab}\right.\\
\left.-\Lambda g_{ab}\left(1+\frac{1}{3}\alpha^{2}\psi^{4}\right)\right]\left(1-\alpha\psi^{2}\right)\\
+\alpha\left[\partial_{a}\partial_{b}\left(\psi^{2}\right)-\varGamma_{ab}^{\mu}\partial_{\mu}\left(\psi^{2}\right)-6\partial_{a}\left(\psi\right)\partial_{b}\left(\psi\right)+\left(\partial\psi\right)^{2}g_{ab}\right]=\\
\kappa\left\{ T_{ab}^{+}\varTheta\left(w\right)+T_{ab}^{-}\varTheta\left(-w\right)+S_{ab}\delta\left(w\right)\right.\\
\left.-\frac{1}{2}g_{ab}\left[T^{+}\varTheta\left(w\right)+T^{-}\varTheta\left(-w\right)+S\delta\left(w\right)\right]\left(1-\frac{1}{3}\alpha\psi^{2}\right)\right\} ,\label{eq:12}
\end{multline}
and
\begin{multline}
\left[-g^{ab}\partial_{w}K_{ab}+Tr\left(K^{2}\right)-\Lambda\left(1+\frac{1}{3}\alpha^{2}\psi^{4}\right)\right]\left(1-\alpha\psi^{2}\right)\\
+\alpha\left[\partial_{w}^{2}\left(\psi^{2}\right)-6\left(\partial_{w}\psi\right)^{2}+\left(\partial\psi\right)^{2}\right]=\\
\kappa\left\{ T_{ww}^{+}\varTheta\left(w\right)+T_{ww}^{-}\varTheta\left(-w\right)\right.\\
\left.-\frac{1}{2}\left[T^{+}\varTheta\left(w\right)+T^{-}\varTheta\left(-w\right)+S\delta\left(w\right)\right]\left(1-\frac{1}{3}\alpha\psi^{2}\right)\right\} \label{eq:13}
\end{multline}
corresponding to Eqs. \eqref{GC1}, \eqref{GC2}, and \eqref{GC3},
respectively. In these latter equations, we have used the fact that
in the immediate neighborhood of the hypersurface, the energy-momentum
tensor of the bulk spacetime and of the hypersurface itself could
collectively be written as
\begin{equation}
T_{\mu\nu}=T_{\mu\nu}^{+}\varTheta\left(w\right)+T_{\mu\nu}^{-}\varTheta\left(-w\right)+S_{\mu\nu}\delta\left(w\right),\label{energy-momentum}
\end{equation}
where $\varTheta$ and $\delta$ denote the Heaviside step function
and the Dirac delta function, respectively, while $S_{\mu\nu}$ is
the induced energy-momentum tensor on the hypersurface. Furthermore,
note that according to the selection of the Gaussian normal coordinates,
we have $g_{wa}=0$, $g_{ww}=1$ and $S_{wa}=S_{ww}=0$, identically. 

Taking integral of Eqs. \eqref{eq:11}, \eqref{eq:12} and \eqref{eq:13}
over a volume in the neighborhood around the hypersurface from $-\epsilon$
to $\epsilon$, and then operating a limit at which $\epsilon\rightarrow0$,
lead us to the boundary conditions we are looking for. In what follows
we assume that $\psi$ is a continuous function across the hypersurface.
This assumption is somehow necessary in order to avoid the emergence
of mathematically non-acceptable expressions such as a multiplication
of two Dirac delta functions. Starting from Eq. \eqref{eq:11}, we
observe that the only total integrand is $\partial_{w}\partial_{a}\left(\psi^{2}\right)$,
which makes it the only term that survives the limiting process. In
other words, operating $\underset{\epsilon\rightarrow0}{\lim}\intop_{-\epsilon}^{\epsilon}dw$
on $\eqref{eq:11}$ yields
\begin{equation}
\underset{\epsilon\rightarrow0}{\lim}\intop_{-\epsilon}^{\epsilon}\alpha\left[\partial_{w}\partial_{a}\left(\psi^{2}\right)\right]dw=\underset{\epsilon\rightarrow0}{\lim}\alpha\left[\partial_{a}\left(\psi^{2}\right)\right]_{-\epsilon}^{\epsilon}=0,
\end{equation}
which immediately reduces to
\begin{equation}
\left[\partial_{a}\left(\psi\right)\right]_{-}^{+}=0.\label{2nd junction condition}
\end{equation}
In the latter equation, $\left[\,\right]_{-}^{+}$ denotes a jump
in the included term across the hypersurface , i.e. $\left[\partial_{a}\left(\psi\right)\right]_{-}^{+}=\left.\partial_{a}\left(\psi\right)\right|_{+0}-\left.\partial_{a}\left(\psi\right)\right|_{-0}$.
This means that not only $\psi$, but also its first tangential derivatives
are continuous across the hypersurface. However, if $\psi$ is only
a function of the normal coordinate $w$, this condition is self-satisfied.
Applying the similar process to Eqs. \eqref{eq:12} and \eqref{eq:13},
and raising one of the indices in the first equation gives rise to
\begin{equation}
\left(1-\alpha\psi^{2}\right)\left[K_{a}^{b}\right]_{-}^{+}=-\kappa\left[S_{a}^{b}-\frac{1}{2}\delta_{a}^{b}S\left(1-\frac{1}{3}\alpha\psi^{2}\right)\right]\label{3rd junction condition}
\end{equation}
and
\begin{multline}
-\left(1-\alpha\psi^{2}\right)\left[K\right]_{-}^{+}+\alpha\left[\partial_{w}\left(\psi^{2}\right)\right]_{-}^{+}=\\
-\kappa\left[\frac{1}{2}S\left(1-\frac{1}{3}\alpha\psi^{2}\right)\right]\label{4th junction condition}
\end{multline}
respectively. Although Eq. \eqref{4th junction condition} looks to
be different from \eqref{3rd junction condition}, in fact it is not.
To show this, it is enough to apply the integration-limiting process
to the conformal equation given in Eq. \eqref{conformal equation}
to see that 
\begin{equation}
\left[\partial_{w}\left(\psi^{2}\right)\right]_{-}^{+}=-\frac{2}{3}\psi^{2}\left[K\right]_{-}^{+}.
\end{equation}
Direct substitution into Eq. \eqref{4th junction condition} simplifies
it to
\begin{equation}
\left[K\right]_{-}^{+}=\frac{1}{2}\kappa S,
\end{equation}
which is nothing but the trace of Eq. \eqref{3rd junction condition}.
It is not hard to see that for $\alpha\rightarrow0$ (or $\psi\rightarrow0$)
the original Israel junction conditions are recovered, as expected.
The results are in full agreement with Eqs. (3.42) and (3.43) of \citep{Aviles1}.
It is worth-mentioning that all the above calculations were done while
we initially had assumed that the first fundamental form (the metric)
is continuous across the shell. This rather intuitive condition, which
guarantees the smoothness of a hypothetical passage through the shell,
is generally accepted as the first junction condition in all thin-shell
formalisms.

\section{\label{section3}Thin-Shell wormhole construction}

The above junction conditions, Eqs. (\ref{2nd junction condition})
and (\ref{3rd junction condition}) could be used to construct a TSW,
provided that we have an exact solution to the action in Eq. (\ref{action}).
Here we consider a de Sitter black hole solution in \citep{Martinez1}
by MTZ, given with the spherically symmetric line element
\begin{equation}
ds^{2}=-f\left(r\right)dt^{2}+f\left(r\right)^{-1}dr^{2}+r^{2}d\Omega^{2},\label{line element}
\end{equation}
with metric function
\begin{equation}
f\left(r\right)=-\left(\frac{r}{l}\right)^{2}+\left(1-\frac{m}{r}\right)^{2},\label{metric function}
\end{equation}
where $d\Omega^{2}$ is the line element of the unit sphere, $l\equiv\sqrt{3/\Lambda}$
is the cosmological length, and $m$ is the gravitational mass which
is related to the scalar field through
\begin{equation}
\psi\left(r\right)=\frac{\alpha^{-1/2}m}{r-m}.\label{scalar field}
\end{equation}
The solution is given only for a positive cosmological constant and
in general has an inner, an event, and a cosmological horizon at
\begin{equation}
r_{i}=\frac{l}{2}\left(-1+\sqrt{1+4m/l}\right),\label{inner horizon}
\end{equation}
\begin{equation}
r_{e}=\frac{l}{2}\left(1-\sqrt{1-4m/l}\right),\label{event horizon}
\end{equation}
and
\begin{equation}
r_{c}=\frac{l}{2}\left(1+\sqrt{1-4m/l}\right),\label{cosmological horizon}
\end{equation}
respectively. Depending on the values of the mass $m$ and the cosmological
length $l$, the solution may exhibit all three, two or only one horizon.
For $m=0$ there is only a single cosmological horizon at $r_{c}=l$.
For $0<m/l<1/4$ all three horizons are real. For $m=l/4$ the event
and the cosmological horizons, $r_{e}$ and $r_{c}$, coincide and
hence we have only two horizons. Nonetheless, for $m>l/4$ the solution
turns into non-black hole for which the inner horizon changes role
to a new cosmological horizon. This last situation is interesting,
as will be seen in the following lines. It is also evident that in
the limit $\Lambda\rightarrow0$ ($l\rightarrow\infty$) the action
in Eq. (\ref{action}) and the metric function in Eq. (\ref{metric function})
coalesce with the action considered and the solution given by Bocharova,
Bronnikov and Melnikov in \citep{Bocharova1} and Bekenstein in \citep{Bekenstein1},
separately, today known as the BBMB solution. Note that the structure
of the metric function in Eq. (\ref{metric function}) is de Sitter
\textendash{} extremal Reissner\textendash Nordström, whose corresponding
TSW has already been studied in pure Einstein gravity \citep{Eid1}.

To construct the TSW, we follow Visser's cut-and-paste recipe \citep{Visser1}.
By cutting two exact exterior copies from the MTZ spacetime such that
the cutting radius $a$ is greater than (any possible) event horizon
and less than the cosmological horizon, we bring them together at
their common timelike hypersurface $\mathcal{H}\coloneqq r_{\pm}-a=0$,
which is now identified as the throat of the wormhole. Note that the
TSW in this fashion will be symmetric. An asymmetric TSW \citep{Forghani1},
however, could be constructed by specifying different cosmological
constants to the two sides, i.e. $\Lambda_{+}\neq\Lambda_{-}$. Nonetheless,
this could not be done by considering different masses for the two
sides since the mass is coupled to the scalar field via Eq. (\ref{scalar field})
and earlier in the derivation of the junction conditions we required
$\psi\left(r\right)$ to be a continuous function across the shell.
Therefore, $\left.\psi_{+}\right|_{r_{+}=a}=\left.\psi_{-}\right|_{r_{-}=a}$
necessarily results in $m_{+}=m_{-}$. 

\begin{figure*}[tp]
\includegraphics[scale=0.25]{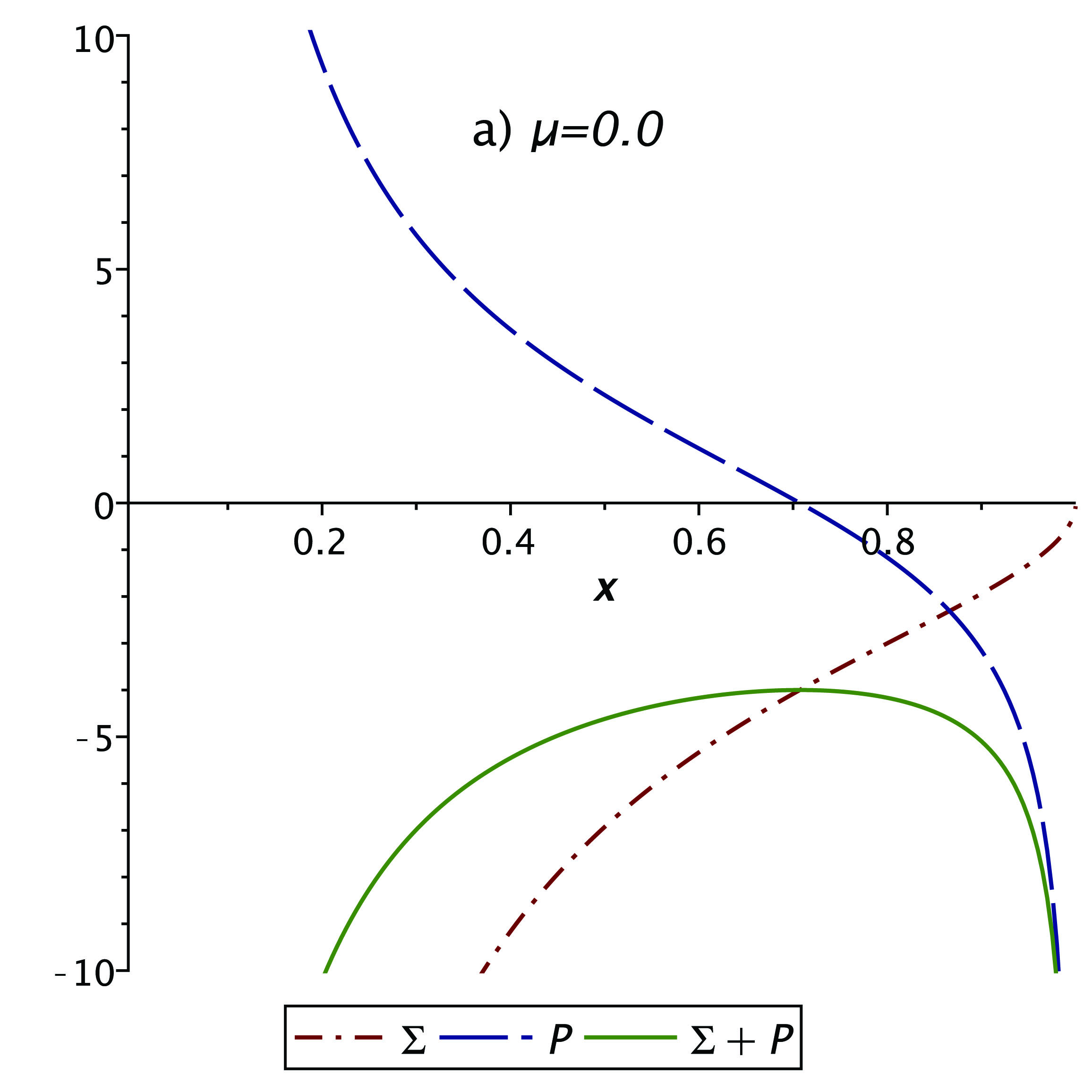}\,\includegraphics[scale=0.25]{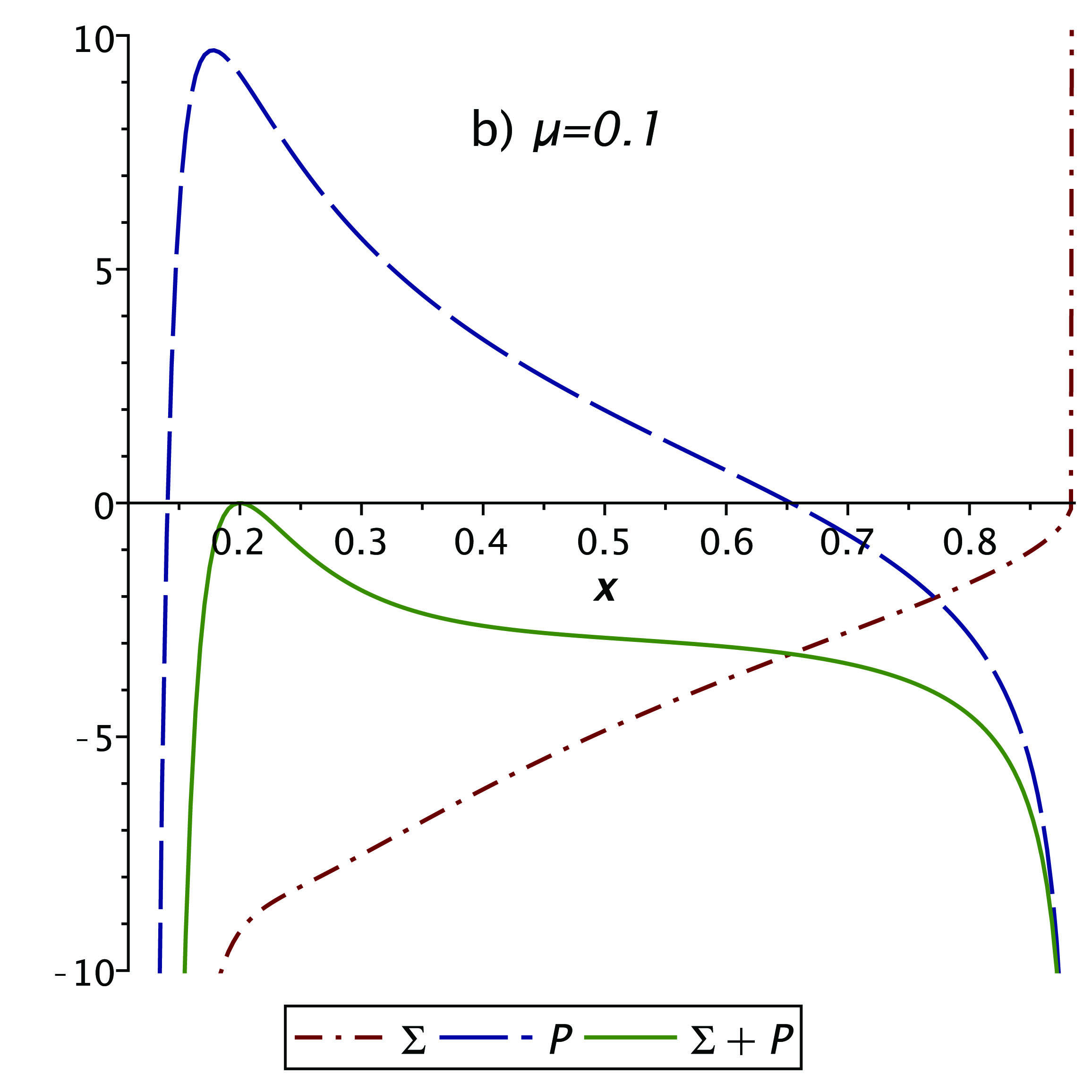}\,\includegraphics[scale=0.25]{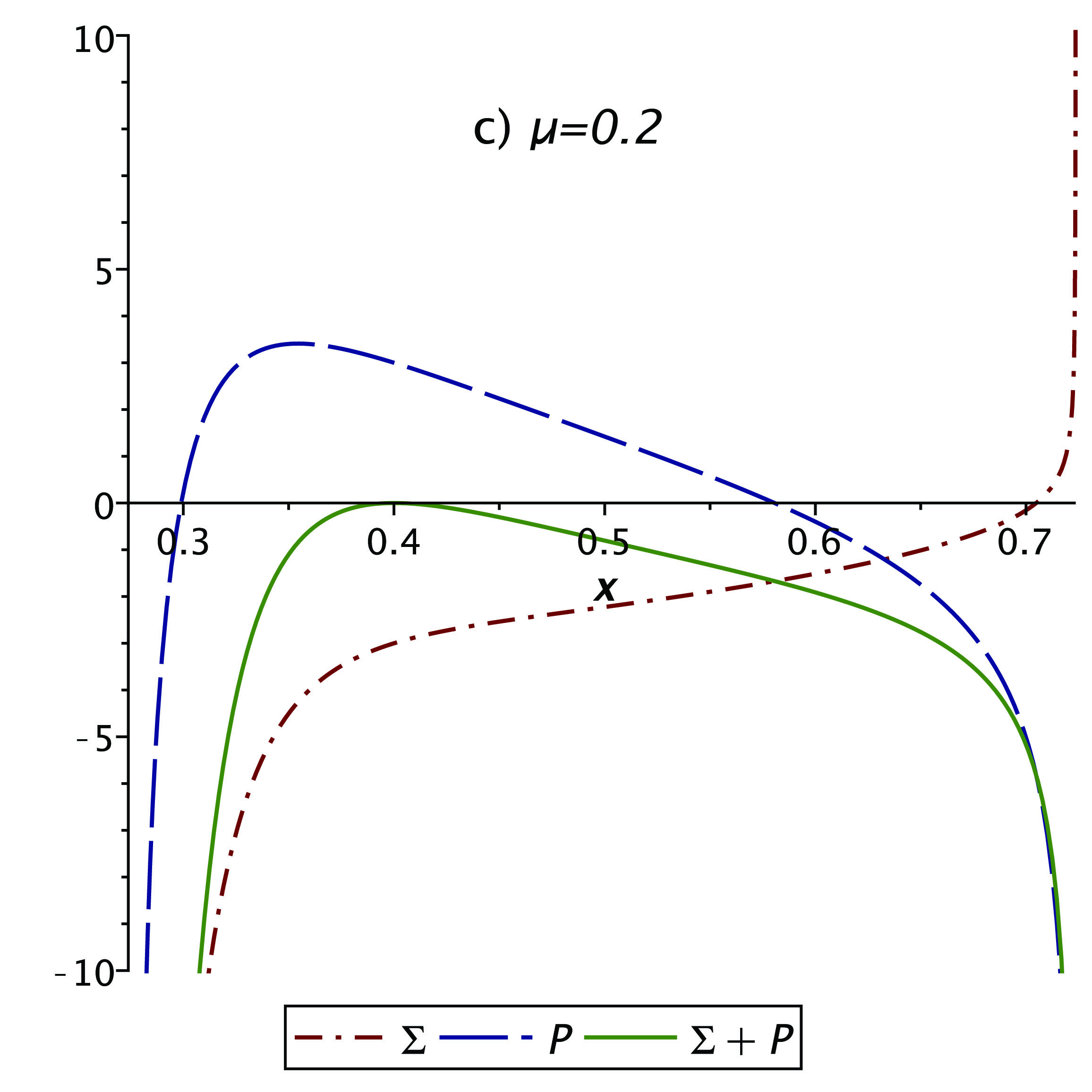}\\
\includegraphics[scale=0.25]{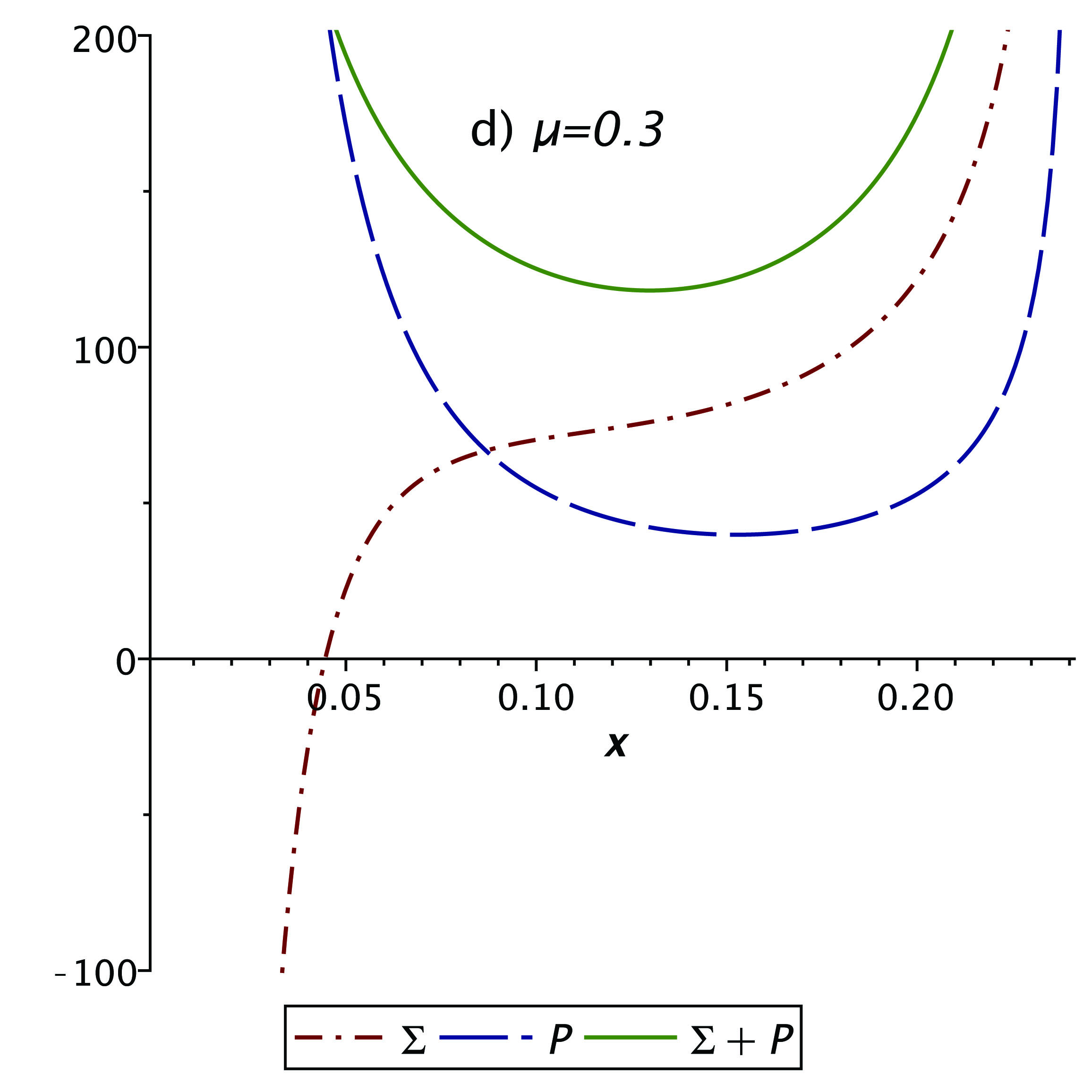}\,\includegraphics[scale=0.25]{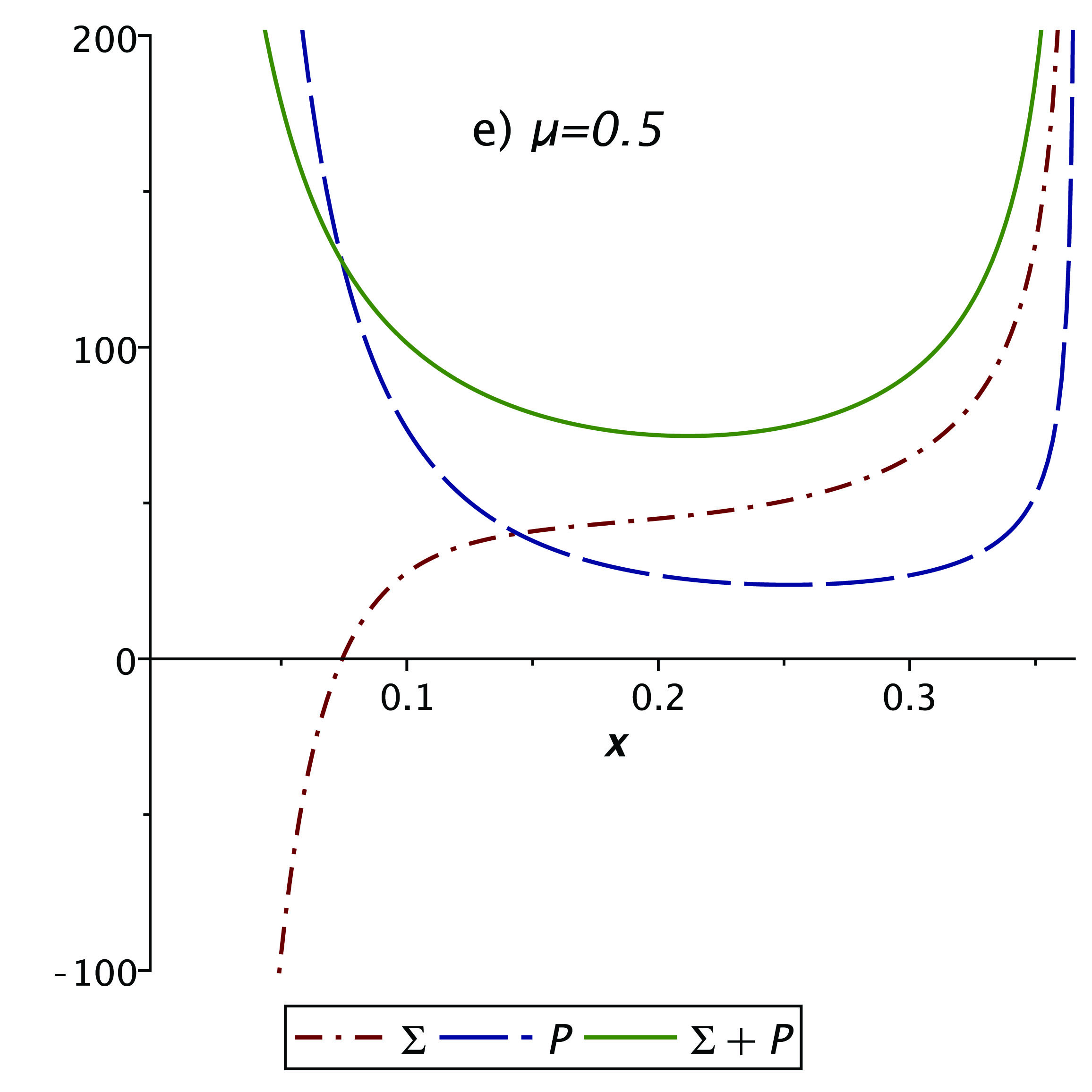}\,\includegraphics[scale=0.25]{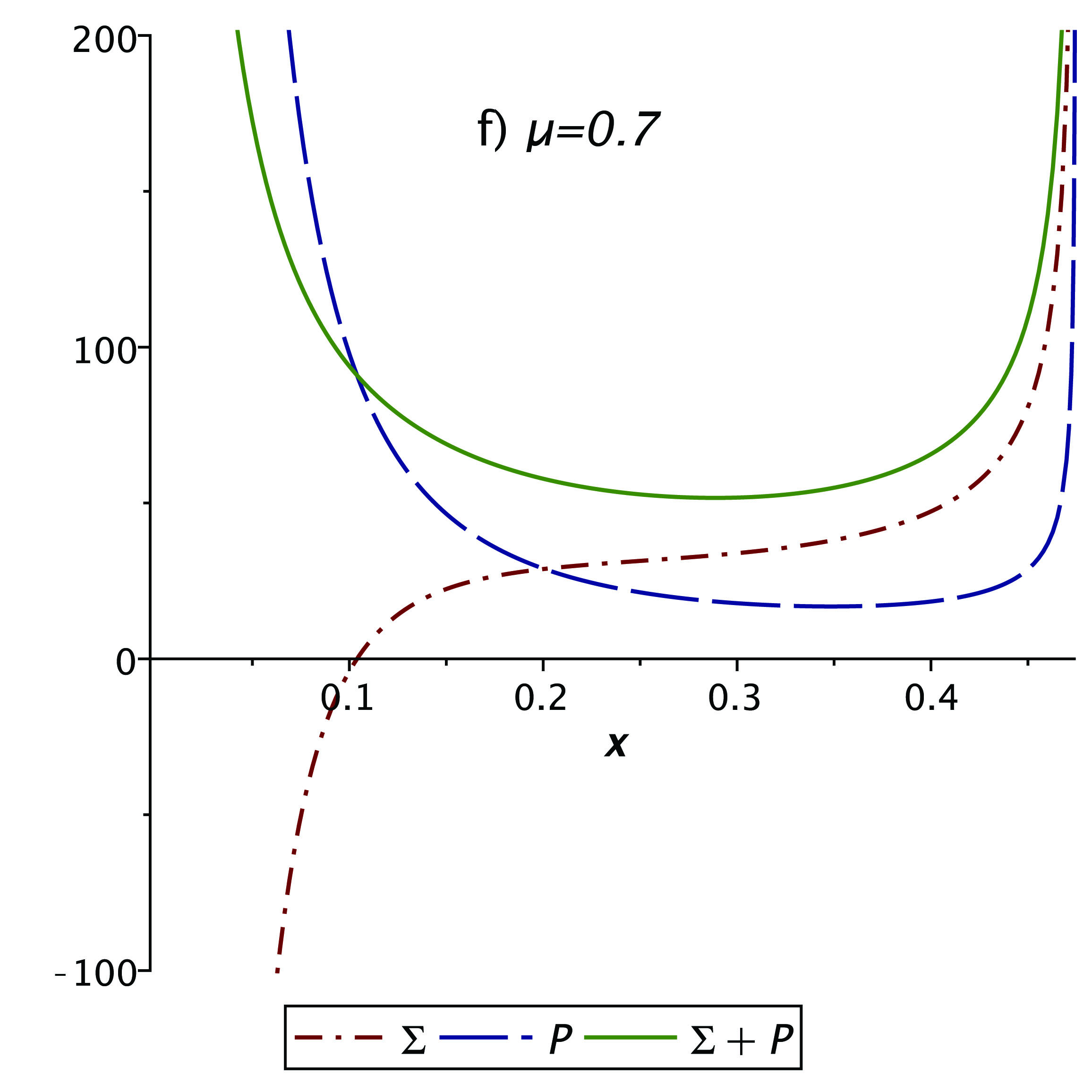}\caption{\label{Fig1}The diagram of the rescaled energy density, $\Sigma\equiv\kappa l\sigma$,
the rescaled pressure $P\equiv\kappa lp$ and their summation $\Sigma+P$,
against the rescaled radius $x\equiv r/l$ for six values of the mass
ratio $\mu\equiv m/l$.}
\end{figure*}

With this symmetric configuration, we will be looking for conditions
under which the TSW could exist without relying on any exotic matter.
The question whether such construction is stable against a radial
perturbation or not will be postponed to future researches. The first
junction condition requires a unique metric for the thin-shell, no
matter which spacetime it is approached from. Therefore, we will have
\begin{equation}
ds_{\mathcal{H}}^{2}=-d\tau^{2}+a^{2}d\Omega^{2}\label{metric of the throat}
\end{equation}
on the shell (on the throat), where $\tau$ and $a$ are the proper
time and the radial coordinate of the shell, respectively. The second
junction condition, given in Eq. (\ref{2nd junction condition}),
is already satisfied since the scalar field is only a function of
the radial coordinate (which happens to be the Gaussian normal coordinate
in this configuration), not the defining coordinates of the throat.
To examine the third junction condition, given in Eq. (\ref{3rd junction condition}),
one needs prior preparations. Firstly, the mixed extrinsic curvature
tensor of the two spacetimes must be calculated. Secondly, a certain
energy-momentum tensor must be considered. Here we employ the energy-momentum
tensor of a perfect fluid which is indicated by
\begin{equation}
S_{a}^{b}=diag(-\sigma,p,p),\label{energy-momentum tensor}
\end{equation}
where $\sigma$ is the energy density, and $p$ is the angular pressure
of the fluid at the throat. To calculate the curvature tensor one
might use a rather explicit definition of it, given by
\begin{equation}
K_{ab}^{\pm}=-n_{\gamma}^{\pm}\left(\frac{\partial^{2}x_{\pm}^{\gamma}}{\partial\xi^{a}\partial\xi^{b}}+\Gamma_{\alpha\beta}^{\gamma\pm}\,\frac{\partial x_{\pm}^{\alpha}}{\partial\xi^{a}}\frac{\partial x_{\pm}^{\beta}}{\partial\xi^{b}}\right),\label{curvature tensor}
\end{equation}
where $x_{\pm}^{\alpha}$ and $\xi^{a}$ are the coordinates of the
bulks and of the throat, respectively, while $\Gamma_{\alpha\beta}^{\gamma\pm}$
are the Christoffel symbols compatible with the bulks' metrics. Furthermore,
$n_{\gamma}^{\pm}$ are the components of the $4$-normal to the throat,
given by
\begin{equation}
n_{\gamma}^{\pm}=\pm\left|g^{\alpha\beta}\,\frac{\partial\mathcal{H}}{\partial x_{\pm}^{\alpha}}\frac{\partial\mathcal{H}}{\partial x_{\pm}^{\beta}}\right|^{-1/2}\frac{\partial\mathcal{H}}{\partial x_{\pm}^{\gamma}}.\label{normal vector}
\end{equation}
Skipping the details of the calculations, the energy density $\sigma$
and the angular pressure $p$ are found to be
\begin{equation}
\sigma=-\frac{2}{\kappa}\sqrt{f}\left[\left(\frac{1}{3}\alpha\psi^{2}\right)\frac{f'}{f}+\left(1-\frac{1}{3}\alpha\psi^{2}\right)\frac{2}{a}\right],\label{energy density}
\end{equation}
and
\begin{equation}
p=\frac{1}{\kappa}\sqrt{f}\left[\left(1-\frac{1}{3}\alpha\psi^{2}\right)\frac{f'}{f}+\left(1+\frac{1}{3}\alpha\psi^{2}\right)\frac{2}{a}\right],\label{angular pressure}
\end{equation}
respectively, where a prime ($'$) denotes a total derivative with
respect to the radial coordinate and all the parameters and functions
are evaluated at the throat's radius. It can be observed that unlike
the forms that generally appear in Einstein's relativity, $\sigma$
is not trivially negative-definite, and hence, there might be a chance
for the weak energy condition (WEC) to be satisfied for the matter
on the throat. To explore this, we directly substitute the metric
function from Eq. (\ref{metric function}), and the scalar field from
Eq. (\ref{scalar field}), into Eqs. (\ref{energy density}) and (\ref{angular pressure})
to obtain
\begin{equation}
\Sigma=\frac{4\left(3x^{5}-3x^{4}\mu-3x^{3}+9x^{2}\mu-8x\mu^{2}+\mu^{3}\right)}{3\sqrt{-x^{4}+\mu^{2}-2\mu x+x^{2}}\left(x-\mu\right)x^{2}},\label{Rescaled energy density}
\end{equation}
and
\begin{equation}
P=\frac{2\left(6x^{4}\mu-6x^{5}-2\mu^{3}+4x\mu^{2}-6x^{2}\mu+3x^{3}\right)}{3\sqrt{-x^{4}+\mu^{2}-2\mu x+x^{2}}x^{2}\left(x-\mu\right)},\label{Rescaled pressure}
\end{equation}
as well as
\begin{equation}
\Sigma+P=\frac{-2\left(x-2\mu\right)^{2}}{\sqrt{-x^{4}+\mu^{2}-2\mu x+x^{2}}x\left(x-\mu\right)},\label{Rescaled sum}
\end{equation}
Herein, we have rescaled the mass by $\mu\equiv m/l$, the radius
by $x\equiv r/l$, the energy density by $\Sigma\equiv\kappa l\sigma$
and the angular pressure by $P\equiv\kappa lp$. Fig. \ref{Fig1}
illustrates the rescaled energy density $\Sigma$, the pressure $P$
and the sum $\Sigma+P$ versus the rescaled radius $x$, for six different
values of $\mu$; $0.0$, $0.1$, $0.2$, $0.3$, $0.5$ and $0.7$.
The radius of the TSW in Figs. \ref{Fig1}a-\ref{Fig1}c , where $\mu<0.25$
($m/l<1/4$), falls between the rescaled event horizon $x_{e}\equiv r_{e}/l=\frac{1}{2}\left(1-\sqrt{1-4\mu}\right)$
and the rescaled cosmological horizon $x_{c}\equiv r_{c}/l=\frac{1}{2}\left(1+\sqrt{1-4\mu}\right)$.
As it is observed from these figures, for $0<\mu<0.25$ although $\Sigma$
becomes positive for radii in the neighborhood of the cosmological
horizon, the sum $\Sigma+P$ is positive nowhere. Therefore, the energy
conditions are not satisfied and the matter is exotic. However, once
$\mu$ exceeds $0.25$ the permissible universe lies within the rescaled
inner radius $x_{i}\equiv r_{i}/l=\frac{1}{2}\left(-1+\sqrt{1+4\mu}\right)$,
which now is the cosmological horizon of this universe. In Figs. \ref{Fig1}d-\ref{Fig1}f,
$\Sigma$ and $P$ have simultaneously become positive for a wide
domain of admissible $x$; and so does $\Sigma+P$. This emphasizes
that, not only the weak energy condition (WEC), but also the dominant
energy condition (DEC) given by $\Sigma>\left|P\right|$, and the
strong energy condition (SEC), given by $\Sigma+P>0$ and $\Sigma+3P>0$,
are satisfied, so the TSW is supported by ordinary matter instead
of exotic.

\section{\label{section4}Conclusion}

For a long time exotic matter violating the energy conditions has
been an indispensable source for the survival of a TSW. In search
for a remedy to this long-standing problem, we resort to new physical
systems that involve new coupling terms. In this study we employed
a black hole solution that involves, in addition to the ERN, a positive
cosmological constant and a self-interacting conformal scalar field
of fourth order. With these new terms, the standard junction conditions
of general relativity for thin-shells must be modified. Accordingly,
we derived the revised form of the junction conditions by using the
Gauss-Codazzi equations. Upon imposing these new junction conditions,
we obtain simple results that are easily tractable. As a result, we
established a new TSW that satisfies WEC and DEC without reference
to exotic sources. To achieve this, however, the mass and the cosmological
radius of the resulting solution must exceed beyond certain minima.

\end{document}